\documentclass[reprint,twocolumn,amsmath,amssymb,aps]{revtex4-1}
\pdfoutput=1 
\usepackage{graphicx}
\usepackage{dcolumn}
\usepackage{bm}
\usepackage{hyperref}
\usepackage{algorithm}
\usepackage{algpseudocode}

\newcommand{\E}{\mathrm{E}}
\newcommand{\Var}{\mathrm{Var}}
\newcommand{\KLG}{\mathrm{KLG}}
\newcommand{\DKL}{\mathrm{D}_\mathrm{KL}}

\begin{document}

\preprint{APS/123-QED}

\title{Monte Carlo Basin Bifurcation Analysis}

\author{Maximilian Gelbrecht}
\email{gelbrecht@pik-potsdam.de}
\affiliation{Potsdam Institute for Climate Impact Research\\ 
Institute of Physics, Humboldt University Berlin}

\author{J\"urgen Kurths}%
\affiliation{Potsdam Institute for Climate Impact Research\\ 
Institute of Physics, Humboldt University Berlin}

\author{Frank Hellmann}
\affiliation{Potsdam Institute for Climate Impact Research
}%

\date{\today}

\begin{abstract}
Many high-dimensional complex systems exhibit an enormously complex landscape of possible asymptotic states. Here, we present a numerical approach geared towards analyzing such systems. It is situated between the classical analysis with macroscopic order parameters and a more thorough, detailed bifurcation analysis. With our machine learning method, based on random sampling and clustering methods, we are able to characterize the different asymptotic states or classes thereof and even their basins of attraction. In order to do this, suitable, easy to compute, statistics of trajectories with randomly generated initial conditions and parameters are clustered by an algorithm such as DBSCAN. Due to its modular and flexible nature, our method has a wide range of possible applications. Typical applications are oscillator networks, but it is not limited only to ordinary differential equation systems, every complex system yielding trajectories, such as maps or agent-based models, can be analyzed, as we show by applying it the Dodds-Watts model, a generalized SIRS-model. A second order Kuramoto model and a Stuart-Landau oscillator network, each exhibiting a complex multistable regime, are shown as well. The method is available to use as a package for the Julia language. 
\end{abstract}

\pacs{Valid PACS appear here}
\maketitle


\section{\label{sec:introduction}Introduction}

Multistability is a universal phenomenon of complex systems. Whether it is hysteresis effects in physics, the human brain \cite{Babloyantz1986, Lytton2008}, gene expression networks \cite{Smolen2000}, in human perception \cite{doi:10.1098/rstb.2011.0254}, power grids \cite{machowski2008} or the climate system \cite{Hirota232, Ciemer2019, May1977, PhysRevLett.122.158701}, almost every sufficiently complex system has a multitude of stable asymptotic states and bifurcations that occur when control parameters are changed. Most traditional methods of bifurcation analysis, such as AUTO \cite{auto} rely on tracking states by continuation of the integration, and become increasingly challenging for high-dimensional systems. Further, for high-dimensional systems, often one is also more broadly interested in classes of asymptotic states such as synchronized versus unsynchronized states of oscillator network or states that share a common symmetry. 
Here, we fill a gap between a coarse analysis with macroscopic order parameters and more thorough bifurcation analysis. 

Our machine learning approach, Monte Carlo Basin Bifurcation Analysis (MCBB), based on random sampling and clustering methods, resolves different classes of asymptotic behaviour into clusters. Rather than studying the existence of states and orbits on the one hand, or only tracking changes in a single order parameter on the other, our approach learns which type of attractors are most dominant in terms of the volume of their basin of attraction, and quantifies the changing size of the basin of attraction of each of these classes as a function of a control parameter. This provides new insights into the bifurcation structure of multistable high-dimensional systems. Thus, we can regard MCBB as a way to interpolate between detailed studies of asymptotic bifurcations tracking every change in asymptotic structure on the one hand, and statistical physics using specialized order parameters to study the macroscopic behavior at the other end.

First we will introduce the method and the idea behind it in the following section. Then, the algorithm will be explained in Sec. \ref{ssec:algo}. A number of paradigmatic examples, the Dodds-Watts model of social and biological contagion and networks of second order Kuramoto oscillators and Stuart-Landau oscillators will follow in Sec. \ref{sec:results}. Lastly, these results and the performance and applicability of the presented method will be discussed in Sec. \ref{sec:discussion}. 

\section{\label{sec:method}Method}

We aim to learn those classes of similar attractors of a high-dimensional system that collectively have the largest basin of attraction with respect to a measure of initial conditions $\rho_0$. Further we intend to understand how they, and their basin volumes, change as a function of a parameter $p$ in a range $I_p$. A class of attractors $\mathcal{C}$ should denote an equivalence class of attractors, including at different $p$, that have similar invariant measures.

To do so we will interpret $\rho_0$ as a probability distribution. We can then draw initial conditions from $\rho_0$ and parameters from $I_p$ and simulating the system to generate trajectories. Assuming ergodicity, the tail of the trajectories then sample the invariant measures on the attractors. We then use these tail samples to estimate whether the invariant measures they were drawn from are similar in the sense of the defining equivalence of our classification. This way we identify clusters among the tail samples that are drawn from the same class. By then computing the number of samples in each cluster drawn at a particular $p$ (or a small interval around it), we provide an estimate for the relative size of the basin of attraction of a class at $p$. Further we can use the samples to study how the members of the class change as $p$ changes.

A key step here is the definition of similarity of invariant measures. Comparing all tail samples to each other is a potentially prohibitively expensive step. Further, in high dimensional systems with a large number of asymptotic states we might be interested in coarser classes of behaviour. Therefore we typically define the similarity between clusters in terms of statistics of the invariant measures that can easily be estimated using the tail samples.

To make this idea more precise we need to define how to determine that two asymptotic measures are similar. We begin by outlining the formal quantities under investigation.

\subsection{Classes of attractors and their basin volumes}

We investigate a complex system with system parameter $p$ yielding a trajectory $\mathbf{x}(t;\mathbf{x}_0, p)$ for initial conditions $\mathbf{x}_0$. This can be an ordinary differential equations system $\dot{\mathbf{x}} = F(\mathbf{x},t;p)$ or a map $\mathbf{x}_{n+1} = F(\mathbf{x}_{n},x_{n-1},...;p)$. If this is a sufficiently well behaved dynamical system, the measure $\rho_0$ will asymptotically evolve into $\rho_{\infty}$, a linear combination of invariant measures on the attractors $\mathcal{A}$ of the system,
\begin{align}
    \rho_\infty = \sum_\mathcal{A} b_\mathcal{A} \rho_\mathcal{A} \; .
\end{align}

As we vary the parameter $p$, the set of attractors and invariant measures of the system will change as well. Given a notion of similarity of invariant measures we define equivalence classes of asymptotic states $\mathcal{C}$. Denoting $\mathcal{C}_p$ those elements of the equivalence class that occur for the system parameter $p$, we have a parameterized space of measures for each class. Assuming that there are only finitely many at each $p$, we write
\begin{align}
\rho_\mathcal{C}(p) = \sum_{\mathcal{A} \in \mathcal{C}_p} c^{\mathcal{A}} \rho_\mathcal{A} \; ,
\end{align}
for asymptotic measures with support only in class $\mathcal{C}$. We assume $\rho_\mathcal{C}(p) = 0$ and  $b_\mathcal{C}(p) = 0$ if the sum is empty and $\sum c^{\mathcal{A}} = 1$ otherwise. Under these assumptions $\rho_\infty$ can be decomposed into classes at each $p$:
\begin{align} 
\rho_\infty (p) = \sum_{\mathcal{C}} b_\mathcal{C}(p)\rho_\mathcal{C}(p)
\end{align}

When we sample from $I_p$ and $\rho_0$m then run the system, the resulting trajectories will have probability $b_\mathcal{C}(p)$ to asymptotically sample an invariant measure in $\mathcal{C}$.

\subsection{Similarity of asymptotic measures}

The key challenge to make this idea operational is to define a notion of similarity. We will approach this challenge to define a computable pseudometric in the following. Let us first consider an extremal case: A linear response of asymptotic measures suggests to identify $\rho_\mathcal{A}(p)$ and $\rho_\mathcal{A}(p+\Delta p)$ as belonging to the same class if they are connected by a smooth continuum of measures. That is, the difference between them vanishes smoothly in an appropriate sense as $\Delta p$ goes to zero, e.g. in the sense of \cite{ruelle1997differentiation, ruelle2009review}. When sampling trajectories, we can build clusters of samples by requiring some discrete notion of this continuity, ensuring that it converges to the right continuum condition in the appropriate limit.

Taking classes built up in this way puts us firmly in the realm of bifurcation analysis. We would resolve every potential difference in asymptotic states. As noted above this might not be desirable when the number of asymptotic states is large, and designing a discrete similarity measure on the high dimensional space that is not prohibitively expensive to evaluate is not straightforward.

Going into the other extreme are order parameters. We could consider $\rho_\mathcal{A}(p)$ and  $\rho_\mathcal{A}(p+\Delta p)$ as similar if they lead to the same order parameter up to some finite bound. This would place us directly into the realm of statistical physics, but requires us to know already what meaningful order parameters for our system are.

Generally speaking we build the classes by making use of some pseudometric on the space of measures built from a weighted sum of differences of statistics $S_k(\rho)$ of the measures. The sampled trajectories then provide us with a way to estimate these statistics, and thus the pseudometric distance between the underlying invariant measures:

\begin{equation}
    D(\rho^i, \rho^j) = \sum_k w_k |S_k(\rho^i) - S_k(\rho^j)|
\end{equation}

Specifically we will show that for the examples considered in this paper it is sufficient to track the mean and the variance of the measures, encoding the position and size of the attractor in phase space:
\begin{itemize}
    \item The position of the attractor: \[\E_k = \langle x\rangle_{\rho_k}\]
    \item The size of the attractor: \[\Var_k = \left\langle (x - E_k)^2 \right\rangle_{\rho_i}\]    
\end{itemize}
where $\rho_k$ denotes the marginal distribution on system dimension $k$.

We further consider the histograms of these statistics over the dimensions of the system. This is particularly useful when the system consists of many identical elements, and it allows us to identify asymptotic states related by permutation symmetry. This is critical for the application to networked systems, for example a dynamical system on a fully connected network will have a symmetry group $S_n$. A more detailed discussion of the technical aspects are given in the next section.

Dependent of the investigated systems, other statistics, such as higher moments or entropy measures can be used as well. Our implementation of the algorithm provides a flexible framework for this purpose (see Appendix \ref{sec:a:julia}).

\begin{figure}
    \centering
    \includegraphics[width=0.4\textwidth]{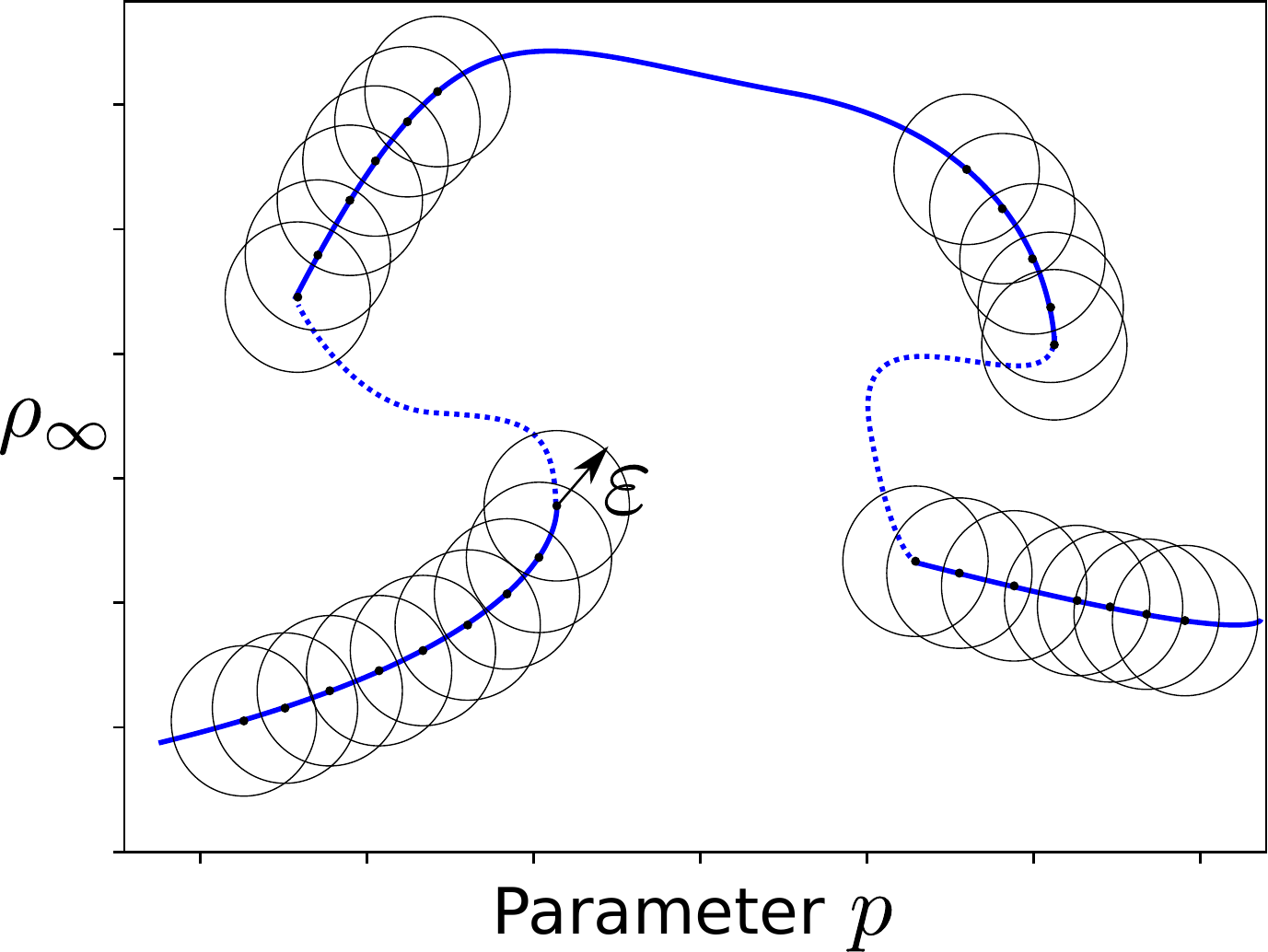}
    \caption{Schematic illustration of an example dynamic with stable asymptotic states (solid blue lines) and unstable asymptotic states (dashed blue lines)}
    \label{fig:epsilon}
\end{figure}

\subsection{Clustering}

Finally to construct a cluster of samples from the estimates of the distance of measures there are two options. Again following linear response theory, we can require that the observed distance is (up to a factor) a finite scaling of the linear response of the asymptotic state. For every sample with a parameter $p^{i}$ we continue the integration with $p^{i}\pm \delta p $ where $\delta p \approx <\min_{j}(||p^{(i)}-p^{(j)}||)>_i$ should be a typical parameter spacing, leading to samples from the measure $\rho^{i\pm}$. Then, we compare the difference $D(\rho^i, \rho^j)$ between trial $i$ and $j$ with the difference to the results of the continuation of the integration $\delta_i^\pm = D(\rho^i,\rho^{i\pm})$. If the former is much larger we assume that there is no direct continuation between the states. Two states are then in the same cluster if there is a chain of states connecting them.

Instead of this computationally intensive continuation study we can also try to extract sensible values for the distance between samples directly from the data. This leads then to a constant response size parameter $\epsilon_{DB}$ for all trials that is ideally a specific percentile $Q_k(p(\delta_i^\pm))$ of the distribution of actual responses $\delta_i^\pm$. When we incorporate the parameter proximity constraint with a weight $w_p$ in the distance calculation as well, the new condition then reads 
\begin{align}
     &i \text{ and } j \text{ are connected if }\nonumber\\ &\sum_k w_k |S_k(\rho^i) - S_k(\rho^j)| + w_p |p^{(i)}-p^{(j)}|< \epsilon_{DB}.
\end{align}
Such a criterion is part of density based clustering algorithms such as DBSCAN  \citep{Ester1996} which we can thus use to distinguish the different classes of asymptotic states given a certain set of suitable statistics. If a single, constant threshold like $\epsilon_{DB}$ is used, it also allows us to vary this threshold in order to resolve different classes of asymptotic finer or coarser: if we choose a large $\epsilon_{DB}$ many similar asymptotic states will be grouped into a single cluster that corresponds to a broad class of asymptotic states. Contrary, a smaller $\epsilon_{DB}$ will result in more different clusters, hence resolving the asymptotic states finer. Fig. \ref{fig:epsilon} schematically illustrates that: As long as this constant threshold is smaller than the minimal distance between trajectories of the two asymptotic states in question, they will be resolved into different clusters. 

Crucially, all steps described here can be performed in a time that scales at most quadratic in the system dimension. This means that high dimensional systems are amenable to being studied in this way.

\subsection{\label{ssec:algo}Algorithm}

We now describe the algorithm that implements the ideas described above in more detail.

MCBB is a modular algorithm: most steps can be modified to suit the dynamical system in question. Algorithm \ref{alg:sbba}  summarizes this procedure and in the following a detailed description of every step is given. 

\paragraph{Setup} We aim to distinguish different classes of asymptotic states by using clustering algorithms on sets of measures that each evaluate one of the $N$ Monte Carlo trials. Given a dynamical system such as an ordinary differential equation system $\mathbf{\dot{x}} = F(\mathbf{x},t; p)$ or a map $\mathbf{x}_{n+1} = F(\mathbf{x}_n, \mathbf{x}_{n-1}, ...; p)$ with $\mathbf{x}\in\mathbb{R}^{N_d}$, we draw $N$ initial conditions $\textbf{x}_0^{(i)}$ from the distribution $\mathcal{U}_{IC}$ and $N$ parameter values $p^{(i)}$ from the distribution $\mathcal{U}_p$. In what follows, we will will use uniform distributions for $\mathcal{U}_{IC}$ and $\mathcal{U}_p$. While we will mostly focus on systems with one parameter dimension, it is in principal also possible to investigate systems with more than one parameter dimension. In particular setups with two varying parameters can provide useful insights into the dynamics of the investigated systems. However, results for systems with three or more parameter dimensions are harder to visualize and will need exponentially more trial runs to create sufficient density in the parameter space. In contrast, just as for basin stability, the number of necessary samples does not scale with the dimension of the space of initial conditions.

\paragraph{Integration} Subsequently, the system is solved for all of the $N$ drawn configurations $(\mathbf{x}_0^{(i)},p^{(i)})$. The integration time has to be set appropriately to the system, so that the asymptotic states are reached. After discarding the transient, the system is integrated for a sufficiently long time. While in theory, this choice of a suitable integration time and transient time is highly non-trivial, in practice, one should have prior knowledge about the time scales of the system. In most situations choosing these times at reasonably large values and checking them for individual trajectories is sufficient. A more sophisticated approach will be discussed in future work. 

The Julia package provided with this paper (see Appendix \ref{sec:a:julia}) uses DifferentialEquations.jl \cite{Rackauskas} to solve ODE systems. The library automatically chooses appropriate solvers, such as Tsit5 or Verner methods \cite{Verner2010,Tsitouras2011}. Even though these methods feature an adaptive step width during integration, we save the trajectories at a constant step width, so that the results of all $N$ trials are saved at the same time steps. We then consider the sample provided by a set final fraction of the trajectory.

\paragraph{Evaluation of the Integration} On each of the tail samples generated this way we evaluate a set of statistics, typically we consider some number $N_s$ of statistics per system dimension $N_d$. These include per default the position and size of the attractor as the mean and standard deviation of the tail sample. Other statistics are possible as well, though. Thus, we obtain $N$ matrices of statistics $\mathbf{S}_i$ each $(N_{d}\times N_s)$ sized with elements $S_{i,kl}$. 

\paragraph{Clustering} For most clustering algorithms a distance matrix between all samples is needed. This $(N\times N)$ distance matrix can computed from the $\mathbf{S}_i$s with two different approaches. First, by calculating
\begin{align}
    D_{ij} = \sum_l^{N_s} w_l \sum_{k}^{N_d} |S_{i,kl} - S_{j,kl}|\nonumber\\ + w_{N_m+1} | p^{(i)} - p^{(j)}|\label{eq:dist-direct}
\end{align}
where each measure can be weighted with a weight $w_i$. The parameter values can be included in the distance metric with weight $w_{N_m+1}$ to ensure that similar asymptotic states with strongly different parameter values are distinguished from each other. The other possibility is to first fit a one dimensional histogram $H_{i,k}$ to each statistic $k$ across all system dimensions. This can be advantageous when symmetric configuration of asymptotic states should not be distinguished which is often the case for networks of identical units. The distance matrix then follows with a suitable histogram distance $\mathcal{D}_H(H_{i,k},H_{j,k})$ such as the 1-Wasserstein metric with
\begin{align}
    D_{ij} = \sum_k^{N_m} w_k \mathcal{D}_H(H_{i,k}, H_{j,k}) +  w_{N_m+1} | p^{(i)} - p^{(j)}|. \label{eq:dist-histo}
\end{align}
When all $H_{i,k}$ for one specific statistic $k$ share the same binning and norm, the 1-Wasserstein metric can be computed very efficiently from the empirical CDF of each histogram. The choice of the weights $\mathbf{w}$ depends on the statistics used and the expected asymptotic states. Generally, a good first guess is to give higher moments such as variance and non-normality measures lower weight than the mean.
Given the distance matrix, a clustering algorithm such as DBSCAN \cite{Ester1996}, is used. DBSCAN classifies all points that can be reached through a common $\epsilon_{DB}$-area as one cluster. Estimating an appropriate $\epsilon_{DB}$ parameter is a non-trivial task and there are different possibilities. In \citep{Ester1996} the authors recommend to use the k-Nearest Neighbour (kNN) distance, more specifically the 4NN distance and use the value of the 4NN distance at the first visual knee in the ordered 4NN distance graph of all data points as $\epsilon_{DB}$. Another, yet similar possibility is to use the median of the cumulative kNN distance, where k is a certain percentage of all points, e.g. 0.5\%. As explained in Sec. \ref{sec:method}, the $\epsilon_{DB}$ can also be estimated by continuing the integration and tracking the response of $\mathbf{D}$. In the examples we have studied, this yields similar values like the more empirical kNN-based methods, but is computationally more expensive. This is why the kNN-based methods are preferred for the estimation of the parameter. Fundamentally there is no "right" choice of $\epsilon_{DB}$, in combination with the choice of distance measures it determines how finely we want to distinguish tail samples. While the choice of statistics and weights determines what aspects we look at, $\epsilon_{DB}$ provides us with an overall resolution that we can vary. As the clustering step at this point is very quick, it is easy to scan a variety of values. We will see an example further down where two clusters that are somewhat similar are no longer resolved as we increase $\epsilon_{DB}$. Density-based clustering algorithms such as DBSCAN are sensible to outliers. Input that is strongly dissimilar to all other data is classified as an outlier. For our purpose, this will typically happen when an explosion of multistability, many different, yet dissimilar, asymptotic states occur.

\paragraph{Evaluation of the Clustering} The clustering algorithm $\mathcal{C}$ thus returns the cluster assignments
\begin{align}
\mathbf{C} = \mathcal{C}(\left\{\mathbf{S}_i\right\}) 
\end{align} 
which map each of of the $N$ trials to one of the $N_C$ clusters with $C_i\in[1,N_C]$ being the number of this cluster for trial $i$. The cluster assignments $\mathbf{C}$ enable us to further analyse the system in question. First of all we can track the size of the basin of each class of asympotic states for changing parameters and thus quantify bifurcations and multistability within the system. This is done by computing the amount of trials within a parameter window $[p_{min};p_{max}]$ and sliding this window over the complete parameter range. For each cluster $\mathcal{C}_i$, thus our estimator for the relative basin volume at parameter $p$, $\hat b_{\mathcal{C}_i}(p)$ is 
\begin{align} 
	\hat b_{\mathcal{C}_i}(p) &= ||{CL}_i^{(p)}|| / \sum_j^{N_C} ||CL_j^{(p)}||\\ 
	CL_i^{(p)} &= \left\{j | (C_j=i)\cap\left( p^{(j)}\in[p_{min};p_{max}]\right)\right\}.
\end{align}
In order to further assess the dynamics of each class of asymptotic sets, the statistics are subdivided into the sets belonging to each of the clusters as well. This way we can track, for example, how the position or size of samples in a cluster change as a function of $p$. Investigating solutions of typical trajectories within each cluster can provide insights as well. In Section \ref{sec:results} examples of such analysis are shown. 

All in all, the two main parameters of the method are the weights $\mathbf{w}$ of the distance calculation and the clustering parameter, in case of DBSCAN $\epsilon_{DB}$. As a default for $\mathbf{w}$, we take $w_{E}=1$, $w_{Var}=0.5$, $w_p=1$. In the Section \ref{sec:results} we will explain in more detail for every system why we chose the weights presented. For the clustering parameter, an estimate with the kNN distance or a response analysis is made and if needed this value is increased (decreased) if one wants to resolve more (fewer) clusters. As for most Monte Carlo methods, the number of trials $N$ should be chosen sufficiently large so that the results are independent from it. A reasonable test is therefore to run the experiment twice: if the results differ qualitatively, one has to increase $N$. 

\begin{algorithm}[H]
  \caption{\label{alg:sbba} Monte Carlo Basin Bifurcation Analysis (MCBB)}
  \begin{algorithmic}[1]
   \State \textbf{Given:} A system $\dot{\mathbf{x}} = F(\mathbf{x}, t; \mathbf{p})$ with system dimension $N_{d}$ (can be an ODE system but also a map) 
   \State \textbf{Given:} A set of $N_{s}$ statistics $\{\mathcal{S}\}$ on the components of trajectories $\mathbb{R}^{N_t}\rightarrow \mathbb{R}$ (e.g. mean and variance)
   \State \textbf{Given:} A distribution $\mathcal{U}_{IC}$ of the initial conditions and parameters $\mathcal{U}_p$
   \Statex
   \For{$i\gets 1, N$} 
        \State Sample $N$ initial conditions $\mathbf{x}_{0}$ from $\rho_0$ and $N$ parameter values  $\mathbf{p}$
        \State Solve system for a long trajectory $\mathbf{x}(t; \mathbf{p})$
        \For{$dim\gets 1,N_{d}$}
            \For{$meas\gets 1,N_{s}$}
                \State compute matrix of statistics $S_{i,dim,s} = \mathcal{S}_i(x_{dim}(t))$ on the tail of the trajectories.
            \EndFor
        \EndFor
   \EndFor
   \State \textbf{Obtained:} $N$ $(N_{d}\times N_{s})$-matrices $\mathbf{S}_i$
   \Statex
   \State Compute $(N\times N)$-sized distance matrix $\mathbf{D}$ of all $\mathbf{S}_i$ to each other.
   \State Density-based clustering (e.g. DBSCAN) of $\mathbf{D}$
   \State Analyze cluster memberships and statistics $\mathcal{S}$ for each cluster dependent on the parameter values $\mathbf{p}$
   \end{algorithmic}
\end{algorithm}

\section{\label{sec:results}Results}

MCBB is a method that can be applied to a wide range of dynamical systems. Both, systems with discrete and with continuous state spaces are possible to investigate, as are systems with discrete and continuous time evolution. Typical applications are networks of oscillators as will be shown in the following, but also discrete agent-based models such as the Dodds-Watts model. Every system that returns a trajectory given an initial condition and parameter can in principal be analyzed with MCBB. In the following the Dodds-Watts model, Kuramoto oscillator networks and Stuart-Landau oscillator networks will be investigated with MCBB. The source code of all these results is available as Jupyter notebooks in the supplementary material. 

\subsection{Dodds-Watts model}
\label{ssec:dodds-watts}

The Dodds-Watts model of social and biological contagion \cite{Dodds2004,Dodds2005} is a generalization of contagion models such as the SIS and SIR model \cite[e.g.]{Murray2002}. Given is a population of $N_I$ individuals that are connected to all other individuals. Each of the individuals is either in the susceptible (S), infected (I) or recovered (R) state and has a memory of doses they received within the last $T$ time steps $D_{t,i}$.  At each time steps each individual $i$ comes into contact with another individual $j$ that is randomly selected from all other individuals. If $j$ is infected, $i$ receive a dose $d$ with exposure probability $p$. The amount of the dose $d$ is drawn from a distribution $f(d)$.  The dose adds to the dose memory $D_{t,i}$ of $i$ at time step $t$ so that $D_{t,i} = \sum_{t-T+1}^{t} d_{t',i}$. If the dose memory of an individual exceeds the dose threshold $d_i^*$, it becomes infected. Latter dose threshold $d_i^*$ is drawn from a distribution $g(d^*)$. As soon as $D_{t,i}$ drops below the threshold, the individual recovers with probability $r$ at each time step. A recovered individual becomes susceptible again with probability $s$. One gets the classic SIS model for example for the configuration $s=1$, $g(d^*)=\delta(d^* - 1)$, $f(d)=\delta(d-1)$, $T=1$ with $p$ and $r$ as free parameters. For more details on the model, see \citep{Dodds2005}. 
For this $N_I$ dimensional model with discrete states $s_{i,t} \in [S, I, R]$ and discrete time $t \in [1,2,.., t_N]$ we directly evaluate the count of susceptible $N(S)$ and infected states $N(I)$ within the time evolution of each individual as measures for the algorithm. As shown by \citep{Dodds2005}, there are several configuration which possess multistable regimes where also a mixed population with $N(I)$ unequal $0$ or $N_I$ can be stable. 

In particular we are investigating the two configurations: (A) with $N_I=1000$, $T=12$, $r=1$, $g(d)=\delta(d-3)$, $s=1$ and (B) with $N_I=1000$  $g(d)=0.075\delta(d-1)+0.4\delta(d-2)+0.525\delta(d-12)$, $T=20$, $r=1$ and $s=1$. The number of initially infected individuals is drawn from a uniform distributed between $0$ and $N_I$. We evolve the system for 1000 time steps from which we regard the first 800 time steps as the transient. Configuration (B) is roughly similar to the SIS model but with a dosage memory of $20$ steps and a dosage threshold distribution so that roughly half of the population is quite resilient against becoming infected. For both configurations $N=5000$ trajectories with random initial conditions and parameter values were computed. As both of the measures are equally important, we choose $w_I=w_S=1$ and $w_p=0$, so that we do not use the parameter value in the distance calculation. The distance $\mathbf{D}$ was constructed using histograms of the statistics as described in Sec. \ref{ssec:algo}. 

\begin{figure}
    \centering
    \includegraphics[width=0.38\textwidth]{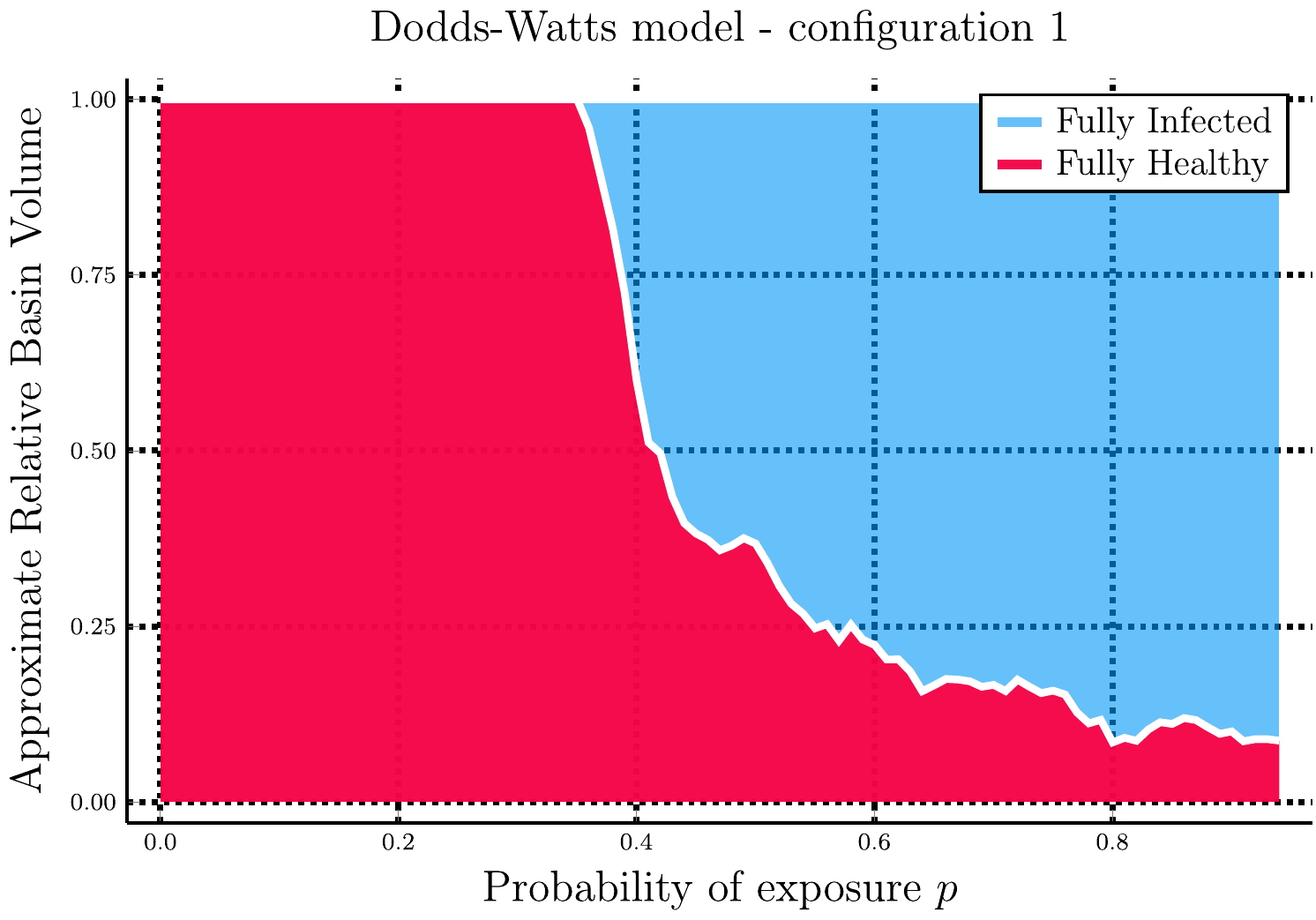}
    \caption{Approximate Relative Basin Volume of the two different classes of asymptotic states, fully infected (blue) and fully healthy (red), for configuration (A) of the Dodds-Watts model. The colored areas in the plot represents the basin volume of the respective state. Computed by using a sliding parameter window over the clustering results (see Sec. \ref{ssec:algo}), a window length of $0.05$ and an offset of $0.01$ were used.}
    \label{fig:dw-1-cm}
\end{figure}

Based on a visual inspection of a 4NN-distance graph, the clustering parameter $d_{DB}=0.15$ was chosen for configuration (A). Fig. \ref{fig:dw-1-cm} shows the results of the analysis. Similar to the results reported in \cite{Dodds2005}, we see for such a configuration a bifurcation occur around $p\approx 0.4$. For values larger than this the fully infected state becomes stable. Its basin of attraction quickly grows, but the fully healthy state remains stable as well with a very small basin of attraction for large $p$ values. 

\begin{figure}
    \centering
    \includegraphics[width=0.95\columnwidth]{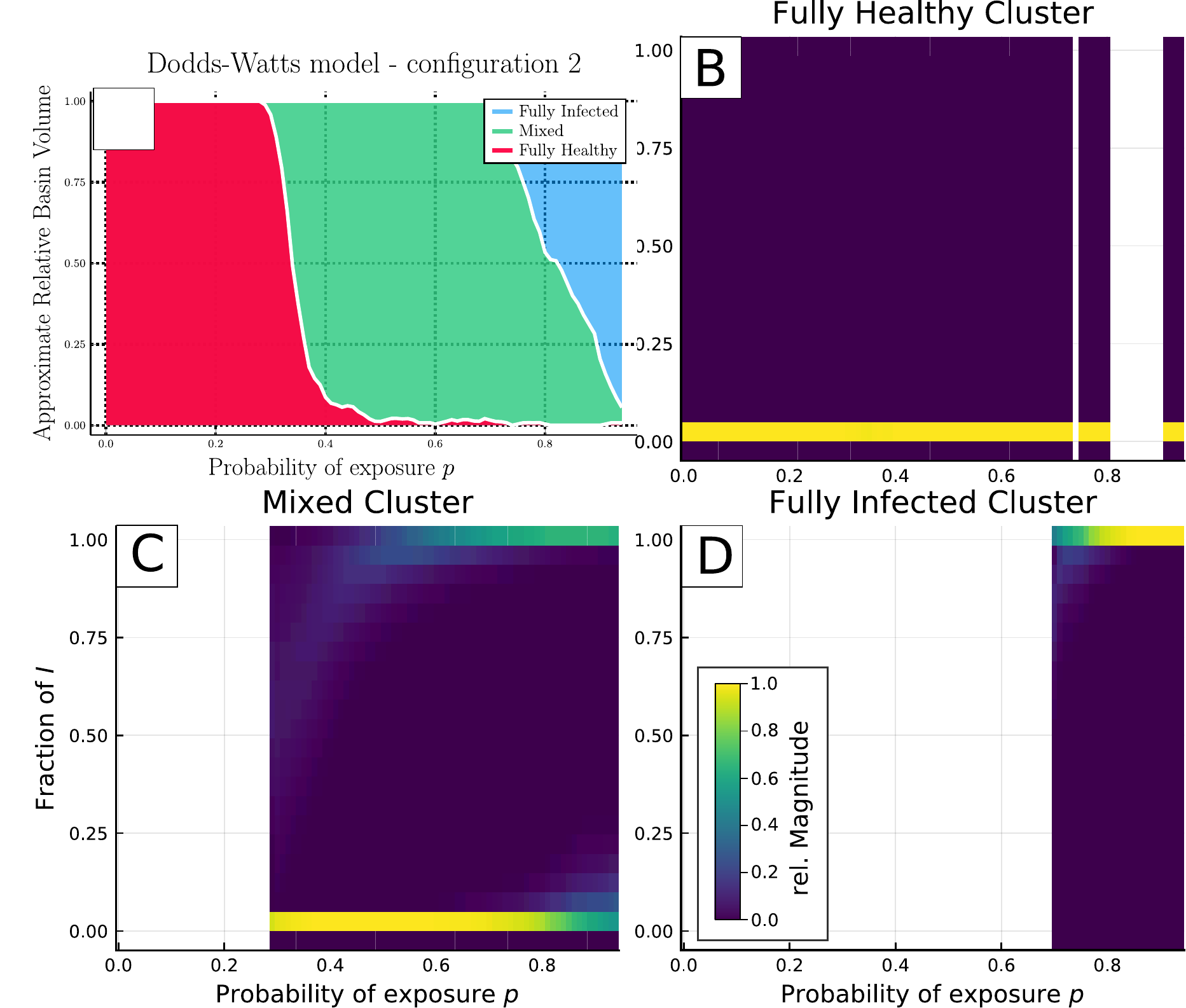}
    \caption{\textbf(A) Approximate Relative Basin Volume of the different asymptotic states of configuration (B) of the Dodds-Watts model. It exhibits a fully infected (blue), fully healthy (red) and mixed state (green). Computed by using a sliding parameter window over the clustering results (see Sec. \ref{ssec:algo}), a window length of $0.05$ and an offset of $0.01$ were used.}
    \label{fig:dw-2}
\end{figure}

Configuration (B) exhibits a slightly more complex structure as Fig. \ref{fig:dw-2} reveals in accordance with the results in \cite{Dodds2005}. Additionally, Fig. \ref{fig:dw-2} features sliding histograms as well. These can be helpful to identify the dynamics of the clusters. For each sliding parameter window a histogram is fitted to all measure results within this window. These histograms are then plotted directly next to each so that we can visualize changes of the measures within each cluster for changing parameter values. In the case of the Dodds-Watts model where we measure the fraction of time an individual agent was infected and susceptible, these are predominantly either $1$ or $0$ as most agents are either infected or susceptible the whole time. Fig. \ref{fig:dw-2}A shows the behaviour of the system. For small values $p$ only the fully healthy state is stable (see also Fig. \ref{fig:dw-2}B). The first bifurcation occurs around $p=0.3$ when a mixed state, for which susceptible and infected individuals coexist, becomes stable. Its basin of attraction quickly grows, while the healthy state remains stable but with a very small basin of attraction. Fig. \ref{fig:dw-2} shows that for growing $p$ the amount of infected individuals rises. Eventually, around $p=0.7$ a fully (or almost fully) infected state becomes stables. As Fig. \ref{fig:dw-2}D shows directly at the bifurcation point not all individuals of the fully infected state are infected which is the case for larger $p$. Comparing the results to these reported in \cite{Dodds2005} we see that the fully infected and the mixed state are indeed two distinctive stable branches of the system and thus rightfully classified by MCBB into two separate clusters. 

\subsection{Kuramoto Networks}
\label{ssec:kuramoto}

The Kuramoto Model is one of the fundamental examples of synchronization theory and network science. The version with inertia has been used in a variety of contexts, most importantly to model nodes in power grids \cite{bergen1981structure, filatrella2008analysis, motter2013spontaneous, menck2014dead, rohden2014impact, hellmann2016survivability}. In the transition towards globally stable synchronization, the Kuramoto model with inertia exhibits an extreme form of multistability, with a large number of attractors. Studying the dominant patterns of synchronization in the transition region was one of the motivating questions for the development of MCBB.

The system is given by the equations
\begin{align}
    \dot{\phi}_i &= \omega_i\, ,\nonumber\\
    \dot{\omega}_i &= \pm 1 - 0.1\, \omega - K \, \sum_j A_{ij} \sin(\phi_i - \phi_j) \,,\label{eq:kura}
\end{align}
with equally many $+1$ and $-1$. For $K=0$ the oscillators rotate freely with $\omega = \pm 10$. As $K$ increases synchronization starts to occur in the network. At $K = 10$ the system typically synchronizes completely with $\omega_i = 0$. While a large number of works have studied the stability of this synchronous state as a function of the local network topology\cite{menck2014dead, rodrigues2016kuramoto,auer2016impact,manik2017network,
schultz2014detours,manik2014supply,hellmann2016survivability,dey2017motif,
kim2016building,schmietendorf2017impact,schafer2017escape,witthaut2016critical}, comparatively little is known about the intermediate regime.

\begin{figure}
    \centering
    \includegraphics[width=0.95\columnwidth]{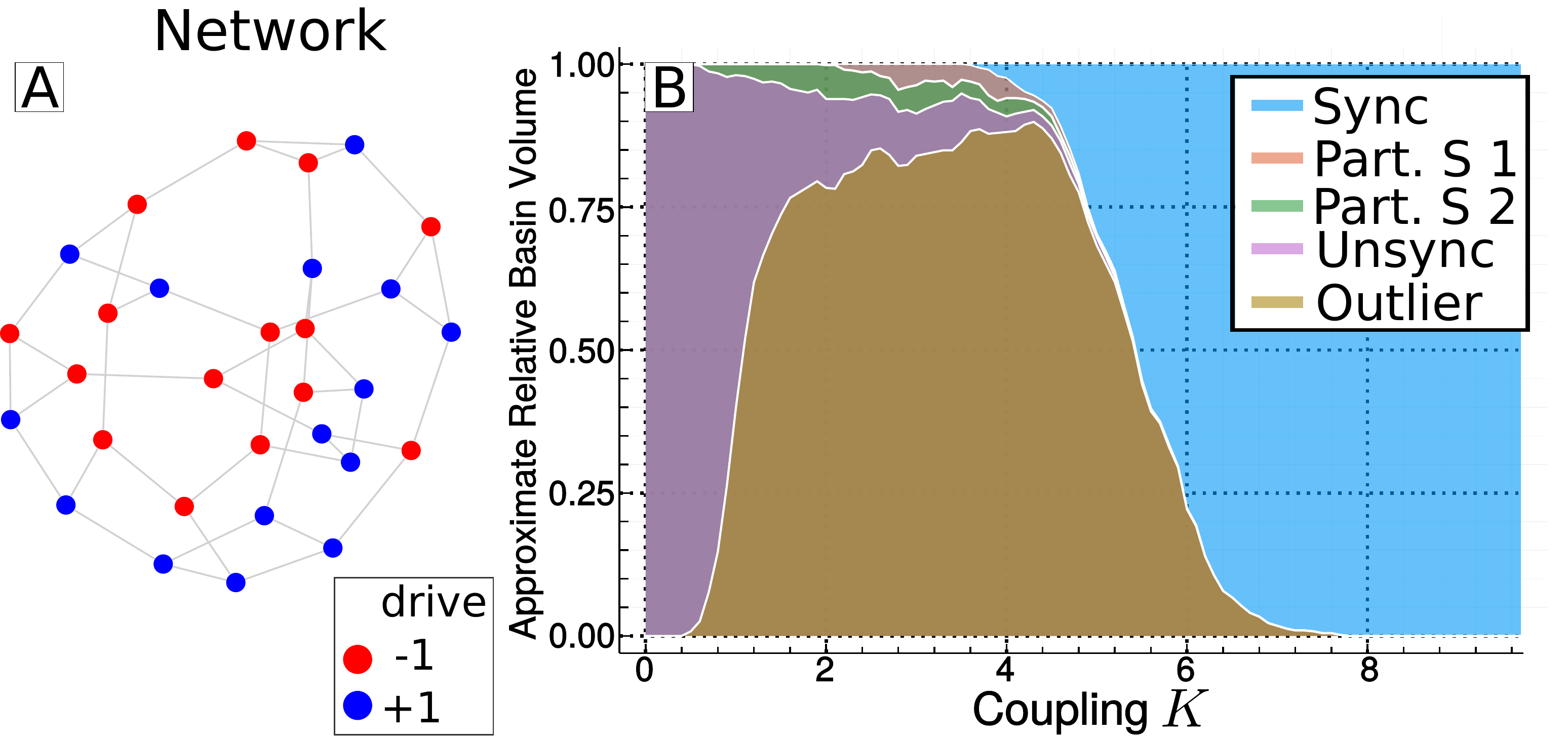}
    \caption{(A) Network structure of the investigated Kuramoto system. Red nodes have a negative drive (minus sign in Eq. \ref{eq:kura}) and blue nodes a positive one. (B) The basin bifurcation diagram for the system with the synchronized, partially synchronized and synchronized regimes.}
    \label{fig:kur-membership}
\end{figure}

As the main dynamics is in the frequency, we will only consider the frequency dimensions in the analysis here. Figure~\ref{fig:kur-membership}a shows the network on which the oscillators are coupled. It is a random regular graph for which every node has degree $k=3$. The statistic we will use on the asymtptotic state are the positions of the frequency of all the nodes and the distance is $\mathbf{D}$ computed according to Eq. \ref{eq:dist-direct}. The results shown are for $N=25,000$ trajectories.

The basin bifurcation structure, with distances calculated from the per-dimension mean of the frequency, is given in Figure~\ref{fig:kur-membership}b. We see that for $K=0$ the oscillators rotate freely, the frequencies are located at $\omega = \pm 10$. This state persists, until its basin starts to shrink from $K=1$ onward. In the intermediate regime most of the asymptotic states occur. These are classed together in the outlier cluster here, meaning that they occur so infrequently that not enough samples can be obtained for a statistical treatment. This shows that the basin structure isn't dominated by one transitional state but an explosion of multi-stability occurs. However, the basin bifurcation diagram also shows two states that achieve a higher basin in the transition region. Each of these clusters occurred in more than 0.5\% of the total runs, and peaks at taking up more than 10\% of the basin volume at some $p$.

\begin{figure}
    \centering
    \includegraphics[width=0.95\columnwidth]{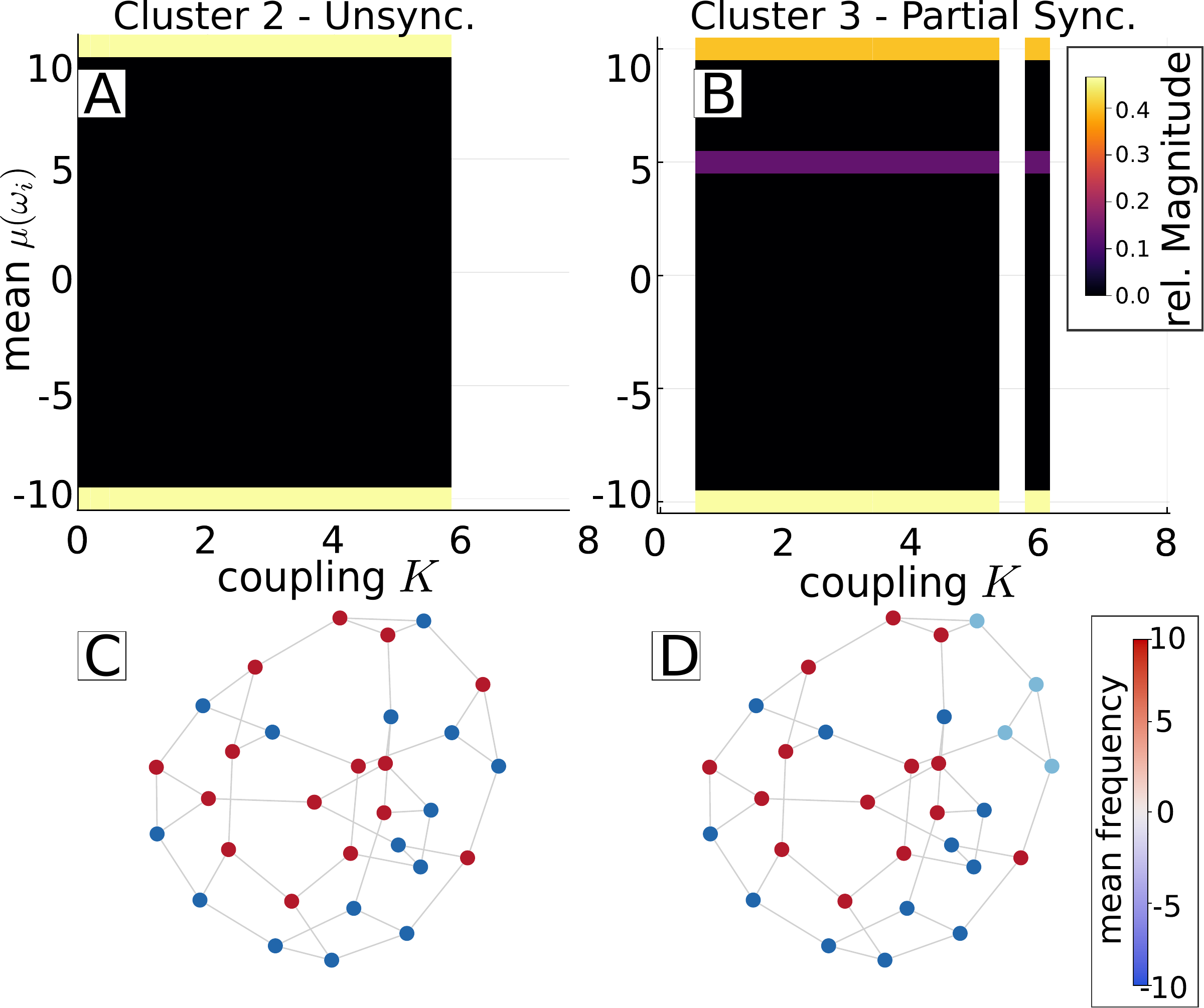}
    \caption{Analysis of the Clusters exhibiting no synchronization (A) and partial synchronization (B). (A) and (B) are sliding histogram plots, similar to Fig. \ref{fig:dw-2} and show the means of the frequencies over all nodes as histograms depending on the coupling parameter. (C) and (D) show the mean frequency of each individual oscillators over all samples in the cluster for cluster 2 (left) and cluster 3 (right).}
    \label{fig:cluster-networks}
\end{figure}

If we look a bit deeper into these clusters, we find that they represent partial synchronization, in which a region of the network is synchronized, while all other oscillators still rotate at their natural frequency Figure~\ref{fig:cluster-networks}.

\begin{figure}
    \centering
    \includegraphics[width=0.95\columnwidth]{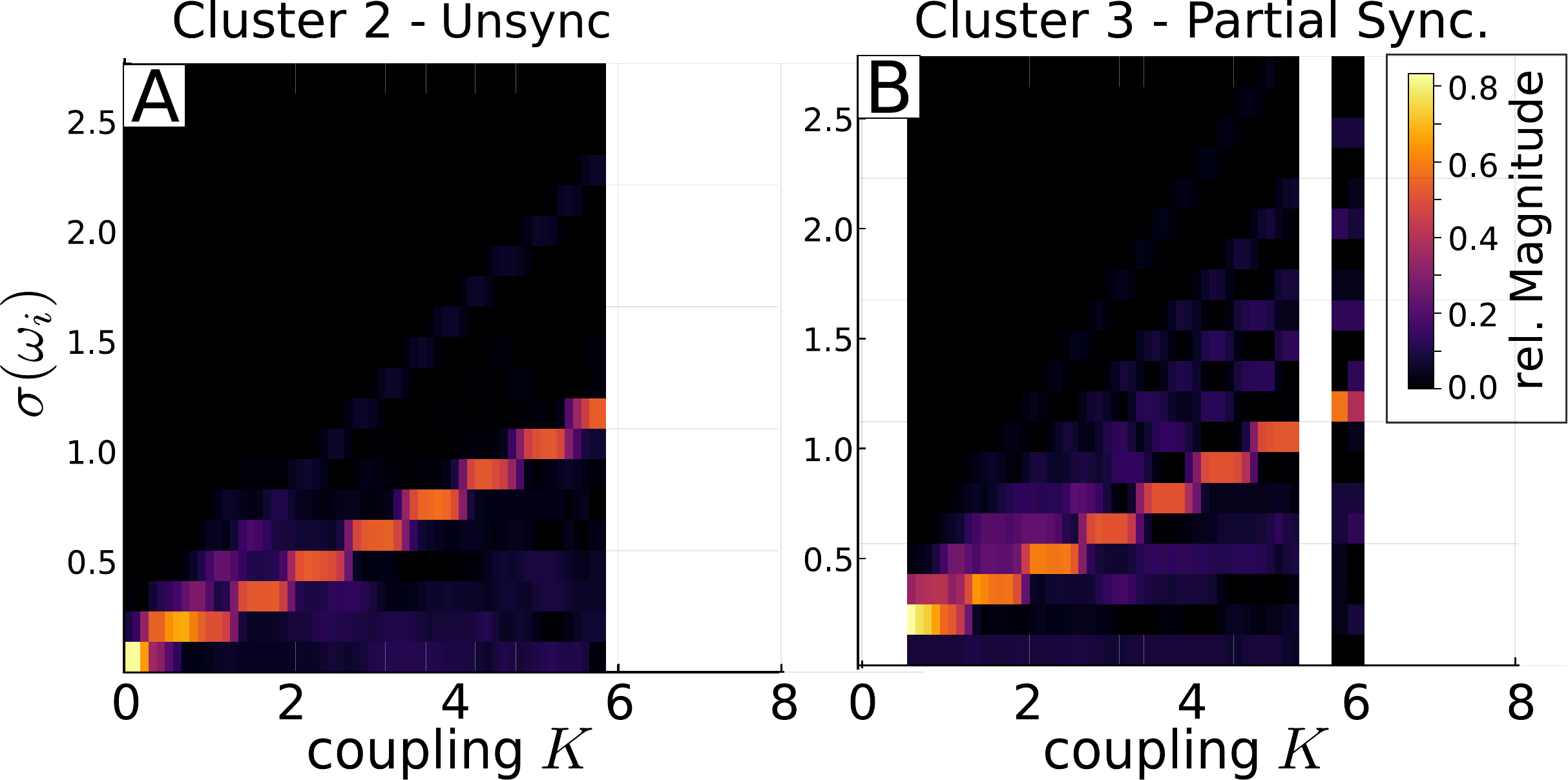}
    \caption{Sliding histogram plot similar to Fig. \ref{fig:cluster-networks}. Here, the standard deviation of all frequency time series is shown depending on the coupling parameter.}
    \label{fig:cluster-histo-var}
\end{figure}

To understand how these intermediate clusters lose stability as $K$ increases, we can consider the size of the asymptotic states considered in Figure~\ref{fig:cluster-histo-var}. Here we see that the size of the attractor increases as $K$ increases. In other words, the frequency itself starts to oscilate around a stable average frequency. This suggests an interesting insight into the behaviour for the transition regime. As $K$ increases some neighbouring oscillators couple and synchronize. As the attractor of the partially synchronized state grows, the oscillators at non-synchronized nodes spend more and more time far from their natural frequency. Eventually they would have to spend considerable time close to the frequency of a synchronized component that they couple to and get entrained.

To verify that these are the mechanisms that drive the transition, and to understand which network properties enable early partial synchronized states, is beyond the scope of MCBB and this paper. However, the basin map of the bifurcation transition that is revealed by this approach provides immediate and crucial insights into how the basin structure and the structure of the attractors themselves change in the transition. In particular it reveals that the attractors do not move, but grow until they lose stability.

\subsection{Stuart-Landau Oscillator Networks}
\label{ssec:sl}

Another paradigmatic type of oscillator is the Stuart-Landau oscillator which can be written as 
\begin{align}
    \dot{z} = (\lambda + i\omega - |z|^2)z
\end{align}
where $z\in\mathbb{C}$, $\lambda$ is the bifurcation parameter and $\omega$ is its eigenfrequency. Originally found by Lev Landau and later derived by Stuart and Watson \cite{landau1944,stuart1960,watson1960} to describe the transition to disturbance in hydrodynamics, it is also a normal form of the Andronov-Hopf bifurcation and hence widely applicable and of great importance in many fields \cite{andronov1971}. Coupling Stuart Landau Oscillator can lead to several interesting phenomena. Most importantly oscillator quenching in the form of Amplitude Death (AD) and Oscillator Death (OD) \cite[e.g]{Koseska2013}. An other interesting phenomena are Chimera states \cite[e.g.]{PhysRevLett.93.174102}. These are states of systems of coupled identical oscillators that exhibit a inhomogeneous pattern in which phase-locked states coexist with drifting states. To apply MCBB for Stuart-Landau systems, we use the configuration of \cite{PhysRevE.98.032301} as it prominently features a multistable regime with travelling wave (TW), oscillation death (OD) and what the authors refer to as stable amplitude chimera (SAC) dynamics. In this setup $N_N$ Stuart-Landau oscillator with identical eigenfrequency $\omega$ are coupled by attractive coupling to its $P_1$ nearest neighbours and repulsive coupling to its $P_2$ nearest neighbours with the following equations:
\begin{align}
    \dot{z}_i = (1 + i\omega - |z_i|^2)z_i + \frac{K}{2P_i}\sum_{k=i-P_1}^{i+P_1} \Re(z_k - z_i)\nonumber\\ - i\frac{K}{2P_2}\sum_{k=i-P_2}^{i+P_2}\Im(z_k - z_i).
\end{align}
where $\Re(x)$ is the real part and $\Im(x)$ the imaginary part of $x$. We can also investigate this setup with the coupling mediated on two Watts-Strogatz random graphs \cite{Watts1998}, one for the repulsive and one for the attractive coupling. With the rewiring probability $p_r=0$, we get the same equation as above, for $p_r\neq0$ we expect changes in the dynamic. 

\subsubsection{Parameter Configuration}
We choose the same parameter configuration as in \cite{PhysRevE.98.032301}: $\omega=2$, $N_N=100$, $P_1=1$ and $P_2=22$. In our experiments we vary $K$, $r_2=P_2/N_N$ and $p_r$. We use random initial conditions with real and imaginary part uniformly distributed between $-1$ and $1$ (in contrast to the cluster initial conditions used for some calculations in \cite{PhysRevE.98.032301}) and vary $K$ from $1.8$ to $2.5$. As per dimension measures we use mean and standard deviation. Since the Stuart-Landau oscillators are complex valued, all measures are applied separately to the real and imaginary part. From our a priori knowledge about Stuart-Landau Oscillators, we know that their asymptotic states will exhibit different kinds of oscillatory behaviour, thus it is a good choice to put the largest weight on the standard deviation. We choose $w_E = 0.25$, $w_{SD}=1$, $w_p=1$ and run $N=15,000$ trials that are integrated from $t_0=0$ to $t_f=200$. The first 70\% of this time span are regarded as the transient and are not used for the evaluation. The first experiment is performed with $p_r=0$ and $r_2=0.22$ and the distance $D$ is calculated using histograms according to Eq. \ref{eq:dist-histo}.

\subsubsection{Varying the coupling}
\begin{figure*}
    \centering
    \includegraphics[width=0.85\textwidth]{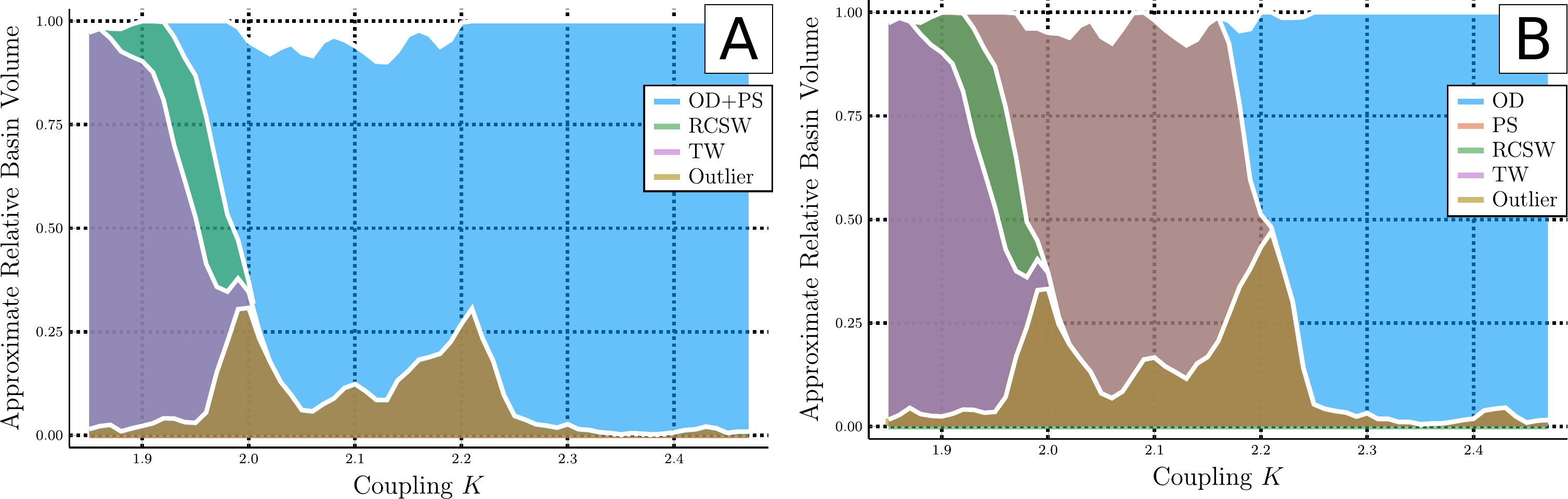}
    \caption{Cluster diagram of the Stuart-Landau Oscillator network with $p_r=0$ for two different values of the clustering parameter $\epsilon_{DB}$. For (A) $\epsilon_{DB}=0.009$ and for (C)\&(D) $\epsilon_{DB}=0.008$. MCBB resolves the different classes of asymptotic states: travelling wave (TW), regular clustered stationary waves (RCSW), (full) oscillation death (OD) and mixed partial synchronized / oscillation death (PS) states. When increasing $\epsilon_{DB}$ states in which most (but not all) oscillators exhibit OD, while the remaining few oscillators are synchronized (PS) and the states in which all oscillators exhibit OD (OD) are merged to one cluster (OD+PS). The window size used is $0.025$ and the offset is $0.01$.}
    \label{fig:sl-1d-cm}
\end{figure*}

\begin{figure}
    \centering
    \includegraphics[width=0.48\textwidth]{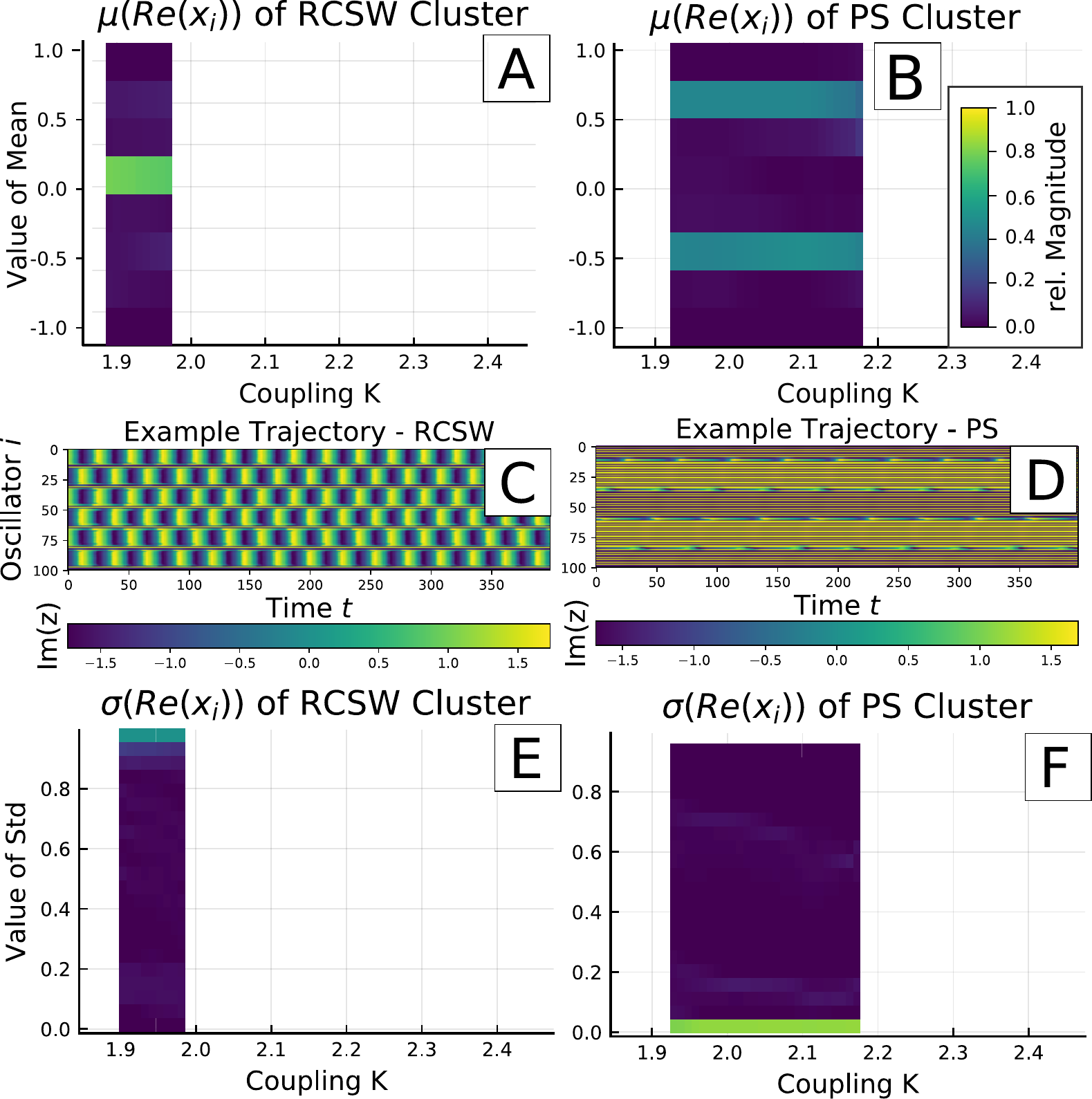}
    \caption{Introspective analysis of the two of the clusters also shown in Fig. \ref{fig:sl-1d-cm}. Plots (A),(B),(E),(F) are sliding histogram plots. For each sliding window of coupling values $K$, the respective measures of trajectories within the said cluster are plotted as a histogram in y-Direction. (A-C) inspect the RCSW cluster. (A),(B) show the mean and the standard deviation of the RSCW cluster. (C) and (D) are example trajectories from the respective clusters. (E), (F) show the mean and standard deviation of the PS cluster.} 
    \label{fig:sl-hists-1d}
\end{figure}

After running the experiment and calculating the distance matrix $\mathbf{D}$, the associated 4-dist graph exhibits the knee point at around $0.01$. We slightly decreased this value to $0.009$ and $0.008$ in the reported results. Figs. \ref{fig:sl-1d-cm} (A) and (B) show these results for the approximate relative basin volume. Similar to the results reported in \cite{PhysRevE.98.032301} we see a multistable regime, in which TW dynamics are prevalent for $K<1.95$ and OD dynamics are for $K>2.2$. In between there are various states in which some oscillators show OD-like behaviour and others exhibit a synchronized oscillation. We thus prefer to refer to these kinds of states as partially synchronized (PS) states. Importantly, the PS states are a mixture of many similarly partially synchronized states and not just a single asymptotic state. If we choose a larger $\epsilon_{DB}$ like in Fig. \ref{fig:sl-1d-cm}A, the states with full OD and the PS states with only few partially synchronized oscillators and otherwise mostly OD dynamics are merged into one cluster (OD+PS). For smaller $\epsilon_{DB}$ they are separated into two distinct clusters (Fig. \ref{fig:sl-1d-cm}A). One particular structured and more common kind of partially synchronized states can be found for $1.9 < K < 2.0$. As Fig. \ref{fig:sl-hists-1d} shows, these states are highly regular stationary waves, interrupted by oscillators exhibiting OD, we thus refer to these states as regularly clustered stationary wave states (RCSW). Aside from these more regular dynamics, there are all kinds of different mixed states between wave-like dynamics and oscillation death. Many are so dissimilar to each other that they fall into the outlier cluster. The outlier cluster has the most members during the transitions from TW to PS via RCSW at $K\approx 2.0$ and at the transition between OD and PS at $K\approx 2.2$. A handful of smaller clusters with less than 60 members (or 0.4\% of all trials) were neglected. They contain partially synchronized states with more similarities to each other than to those in the outlier cluster.
We identified these dynamics by further analyzing the statistics within each cluster. Fig. \ref{fig:sl-hists-1d} shows example plots and sliding histogram plots for two of these clusters. The RCSW states mostly oscillate and thus almost all oscillators have a mean of zero and a constant standard deviation different from zero. We see that these histograms change little for different coupling values. The cluster is very homogeneous with almost all members looking like the example shown in Fig. \ref{fig:sl-hists-1d} C. The PS cluster, on the other hand, is much more inhomogenous. Its members have in common that most of the oscillators exhibit OD, thus as Fig. \ref{fig:sl-hists-1d} confirms, they exhibit nonzero means, with both positive and negative values while having a vanishing standard deviation which corresponds to the typical stable fixpoints of OD dynamics. Fig.\ref{fig:sl-hists-1d} D shows one example, the amount of oscillators still exhibiting a synchronized oscillators is different within the cluster, though. Additional results for the other clusters can be found in the appendix. 

\subsubsection{Varying the coupling and amount of coupled neighbours}
\begin{figure}
    \centering
    \includegraphics[width=0.48\textwidth]{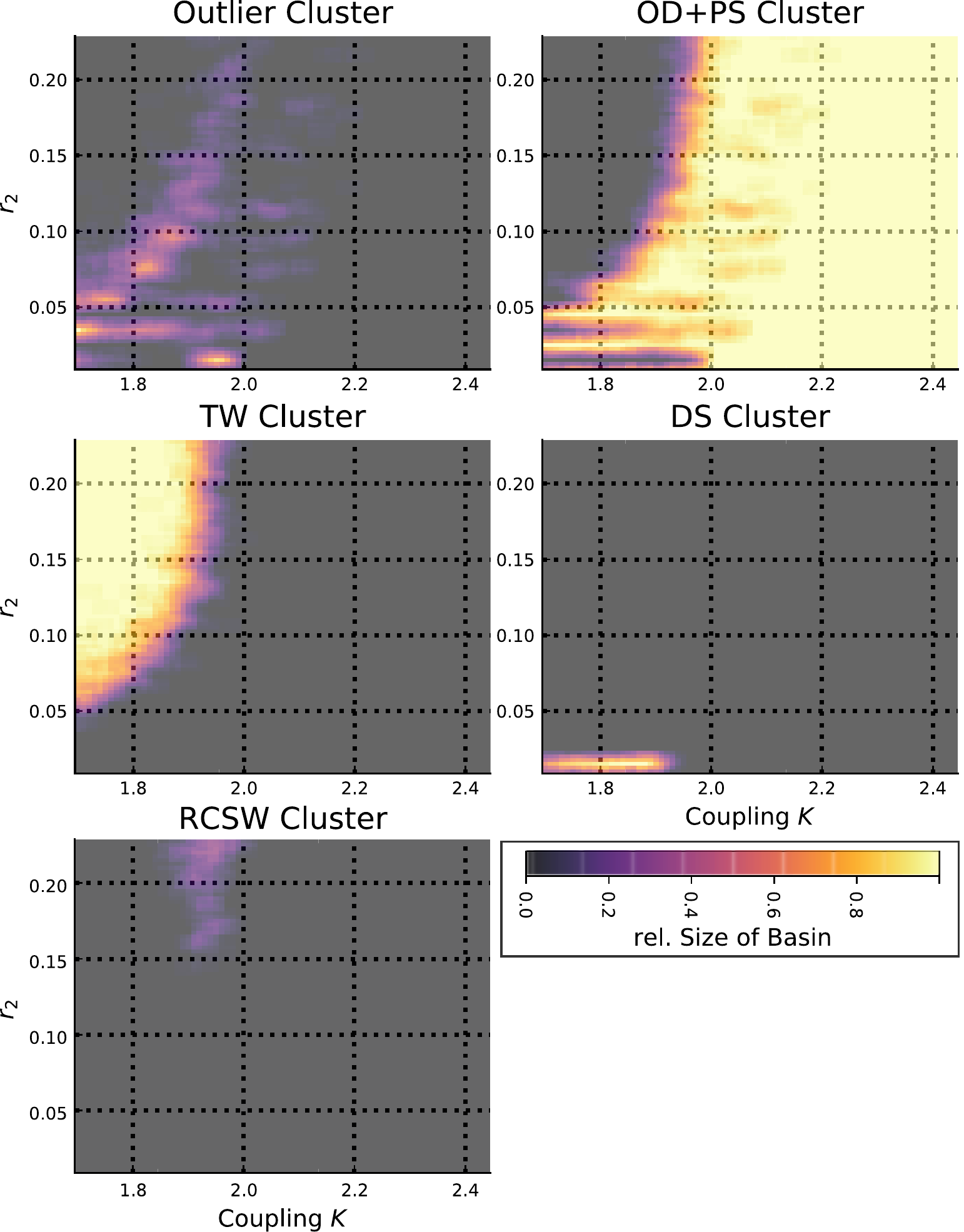}
    \caption{Results from the setup with two parameters, varying the amount of coupled neighbours $r_2$ and the coupling strength $K$. } 
    \label{fig:sl-2d-r2-cm}
\end{figure}

Similarly to the additional setup in \cite{PhysRevE.98.032301}, we can also investigate this system with two varying parameters with MCBB. First, we choose to vary $K$, the coupling, and $r_2$, the relative amount of neighbours the oscillators are coupled to repulsively. Fig. \ref{fig:sl-2d-r2-cm} shows similar clusters of similar asymptotic behaviour as in the one-dimensional setup. We see that TW dynamics are present only for small $K$ and large $r_2$ values, while OD+PS dynamics are present even for small $K$ values when $r_2$ is small. For very small $r_2$ there is also a desynchronized (DS) cluster. Most notably the distinctive RCSW type dynamics are only present for $r_2 > 0.1$ and its basin becomes larger for larger $r_2$ values. 

\subsubsection{Rewiring of the network}
\begin{figure}
    \centering
    \includegraphics[width=0.48\textwidth]{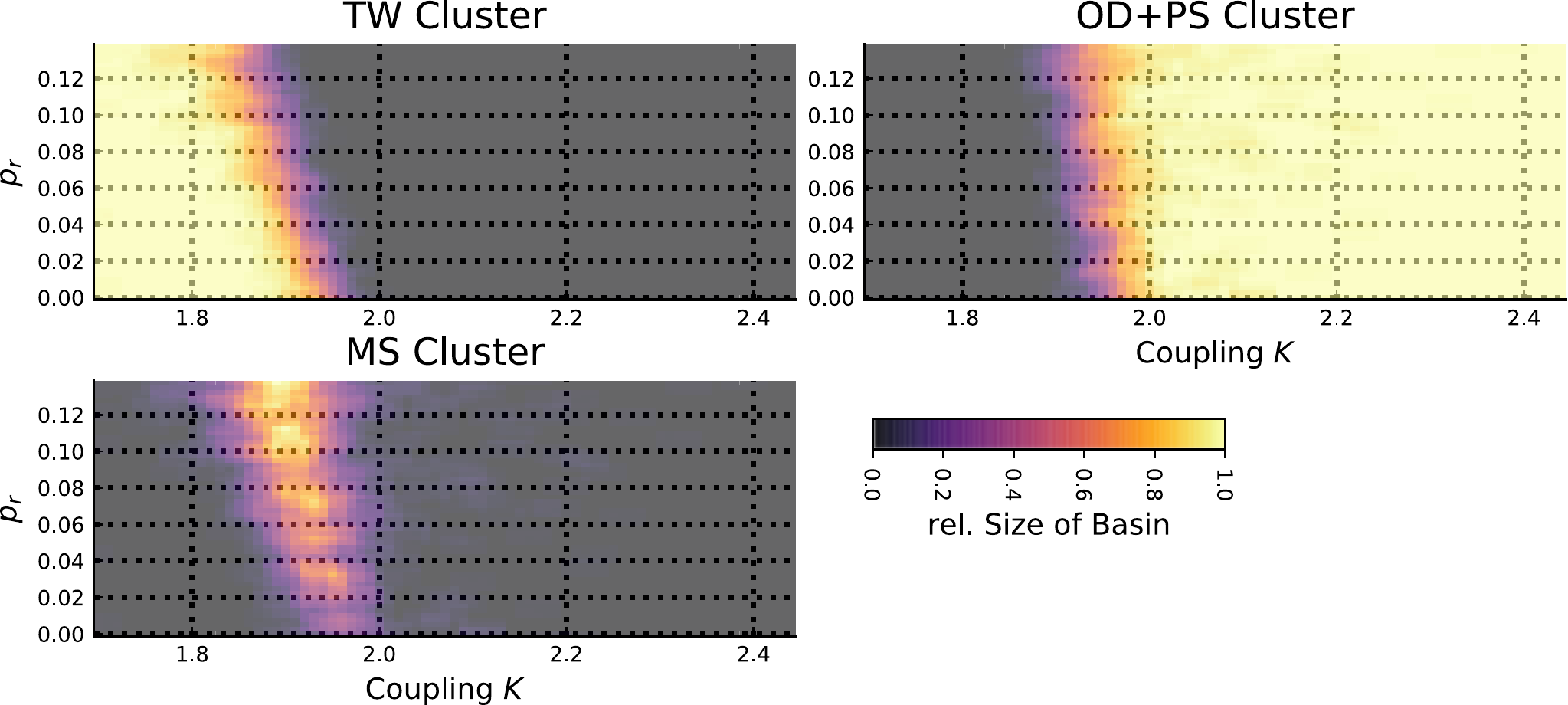}
    \caption{Results from the setup with two parameters, varying rewiring $p_r$ of the Watts-Strogatz random graph that mediates the repulsive coupling and the coupling strength $K$. } 
    \label{fig:sl-2d-pr-cm}
\end{figure}

14 
When we start to randomize the coupling by rewiring it according to the scheme of Watts-Strogatz random graphs, we get the results presented in Fig. \ref{fig:sl-2d-pr-cm}. Here, we added the outlier cluster together with several smaller clusters that all exhibit mixed, partially synchronized, partially OD dynamics to the mixed states (MS) cluster. The range of $K$ for which these kinds of dynamics appear gets wider when the rewiring $p_r$ increases. TW dynamics appear less for larger $p_r$ values. RSCW type dynamics do not appear when we rewire the network.  

\section{\label{sec:discussion}Discussion}

Given a complex system, such as a ODE system, like the Kuramoto and Stuart-Landau networks demonstrated in Sec. \ref{ssec:kuramoto} and \ref{ssec:sl}, or a map like the Dodds-Watts model presented in Sec. \ref{ssec:dodds-watts}, MCBB is able to analyze and quantify which classes of asymptotic states are occuring. As demonstrated with the paradigmatic example systems MCBB is a widely applicable approach. It is suitable to analyze the behaviour of every high-dimensional system that returns a trajectory, be it agent-based models such as the Dodds-Watts model or Differential Equations like the Kuramoto and Stuart-Landau networks. The known bifurcations of these systems were reproduced by MCBB as shown for example with the Dodds-Watts model. Additionally, it enables us to reveal clusters of qualitatively similar asymptotic states for all these systems as the results investigated in Sec. \ref{ssec:sl} show. It does successfully identify the sizes of the basins of the most important asymptotic states even in transition regimes, what a traditional bifurcation analysis can not reveal. For the Kuramoto system we see how and when the basins of the unsynchronized states shrinks and how the basins of the completely synchronized states emerges. We also get an insight into the transition between these states, as we can see how the size of the states increases before they destabilize. Hence, for the Kuramoto model it provides an intuitive way of visualizing the synchronization process. When applying MCBB to a Stuart-Landau system the different asymptotic behaviours, travelling wave states, oscillator quenching phenomena such as oscillator death and mixed stated, are classified in different clusters and interesting dynamics such as regularly clustered stationary wave states are revealed and their basins quantified. 

The analysis can always be fine tuned by changing the clustering parameters to resolve the asymptotic states finer or coarser. Additionally, the weights of the distance calculation provide another mean of adjustment. The flexible nature of the method also allows for experimentation with the statistics used to evaluate the trajectories and the exact clustering algorithm. In particular various entropy-based seem promising to use. While designing the method we already used the per dimension Kullback-Leibler divergence of the time series to the Gaussian measure $\KLG_i = \DKL\left(\rho^G(E_i, \Var_i) \middle\| \rho_i\right)$ as a statistic to track structural changes of investigated systems. This was especially useful for relatively low-dimensional systems. The curve entropy \cite{balestrino2007} of the complete trajectory was tested as well. Additionally, we also experimented with a distance between histograms of the covariance matrices as a statistic. This expands variance-based size measure to also take cross-correlations between the dimensions into consideration which could be useful for systems that exhibit multiple possible cross-correlations structures in the asymptotic states that otherwise behave similar, e.g. different kinds of collective oscillations. For the example systems presented here, it was however sufficient to only use the position and size of the attractors as measures. Additional measures were not necessary to resolve the different classes of asymptotic states. This should not stop experimentation with additional measures though, as some of them are already implemented in the accompanying software as well with further additional ones easy to add.

Aside from the approximate basin volume and the sliding histograms shown in this paper, it is also possible to further investigate the clusters found by the clustering algorithm, e.g. by analyzing which kind of initial conditions lead to certain class of asymptotic states or by analyzing how each dimension is changing with the control parameters separately and not in histogram form. These options are already implemented in the Julia package (see Appendix \ref{sec:a:julia}) and more could be envisioned in the future. 

It is further possible to extend the method to systems with unknown background parameters that adhere to certain distribution and additional control parameters or forcings, such as some climate models which will be further discussed in future work.

While this work focused on introducing the method and testing it with fairly theoretical models, we believe that this opens the door to studying a wide variety of systems in novel ways. We expect that the method will be fruitful in diverse contexts where a mix of multistability and high dimensional behaviour are important. Most notable among those would be biological networks and climate systems.

A distinct limit of the approach is that it is only able to detect and track stable solutions of the investigated systems. Unstable solutions are not accessible with MCBB. A further important avenue of investigation is to study the mathematical properties of the algorithm described here in much more detail. In particular it would be highly desirable to understand the convergence properties of the algorithm. We also suspect that there is considerable scope for improving the clustering by making use of information from the continuation, rather than reverting to a standard density based algorithm. One other avenue of investigation where we will improve the method further is to use the statistics of the tail sample we record in order to track when the integration has reached the asymptotic regime in a suitable sense.

MCBB provides an excellent way to visualize the complex behaviour of systems where a traditional bifurcation analysis is often not useful or difficult to implement. It resolves the most important classes of asymptotic states and enables the user to track the size of its basins along changing parameters.
\appendix

\section{\label{sec:a:julia}Julia Package}

The algorithm is implemented in Julia. It can be installed directly from the GitHub repository \url{https://github.com/maximilian-gelbrecht/MCBB.jl/}. This library makes heavy use of Julia's DifferentialEquations.jl library \cite{Rackauskas}. There is an extensive documentation available that explains the package with many examples that is linked in the page of the repository.   

\section{Logistic Map}

\begin{figure}
    \centering
    \includegraphics[width=0.4\textwidth]{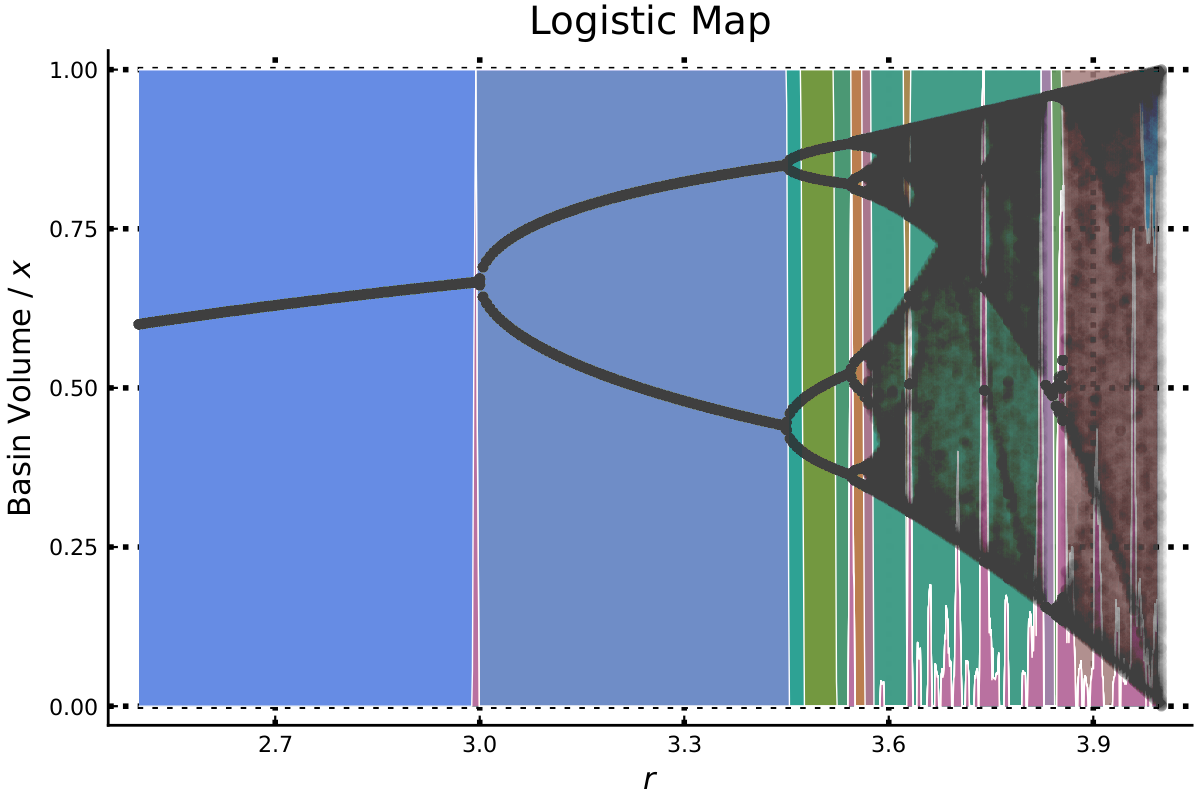}
    \caption{Basin Volume and Bifurcation diagram of a logistic map.} 
    \label{fig:app-logistic}
\end{figure}

While designed for high-dimensional systems, MCBB will also still work in the fringe case of a one dimensional system such as the logistic map $x_{n+1} = r x_n (1 - x_n)$. Fig. \ref{fig:app-logistic} shows the approximate relative basin volume computed with MCBB compared to the bifurcation diagram. It was computed using the mean, standard deviation and Kullbach-Leibler divergence as measures with the weights $1$, $0.5$ and $0.5$. The major bifurcation points are reproduced as do the stable regions inside the chaotic regime form seperate clusters, while most of the chaotic regime is grouped into to distinct clusters, one before and one after the larger stable region around $r\approx 3.8$. 

\section{\label{sec:a:results}More Results}

\begin{figure}
    \centering
    \includegraphics[width=0.45\textwidth]{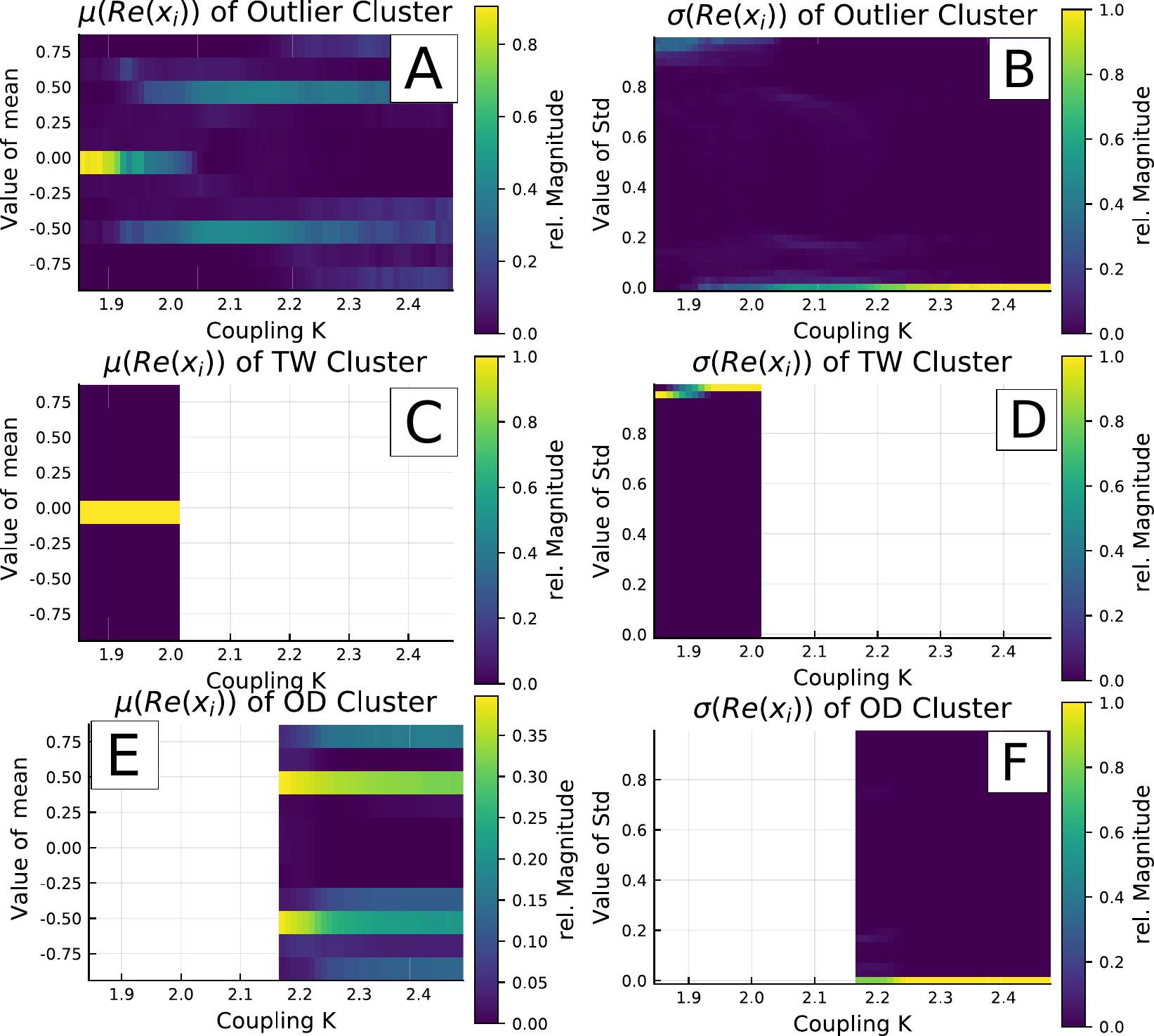}
    \caption{Further analysis on the clusters also shown in Fig. \ref{fig:sl-1d-cm}. (A),(B),(E),(F) are sliding window histograms fits of the denoted measures for trials with parameters within the respective window. (C) and (D) are example trajectories of trials within these clusters.} 
    \label{fig:app-sl-1}
\end{figure}

\begin{figure}
    \centering
    \includegraphics[width=0.45\textwidth]{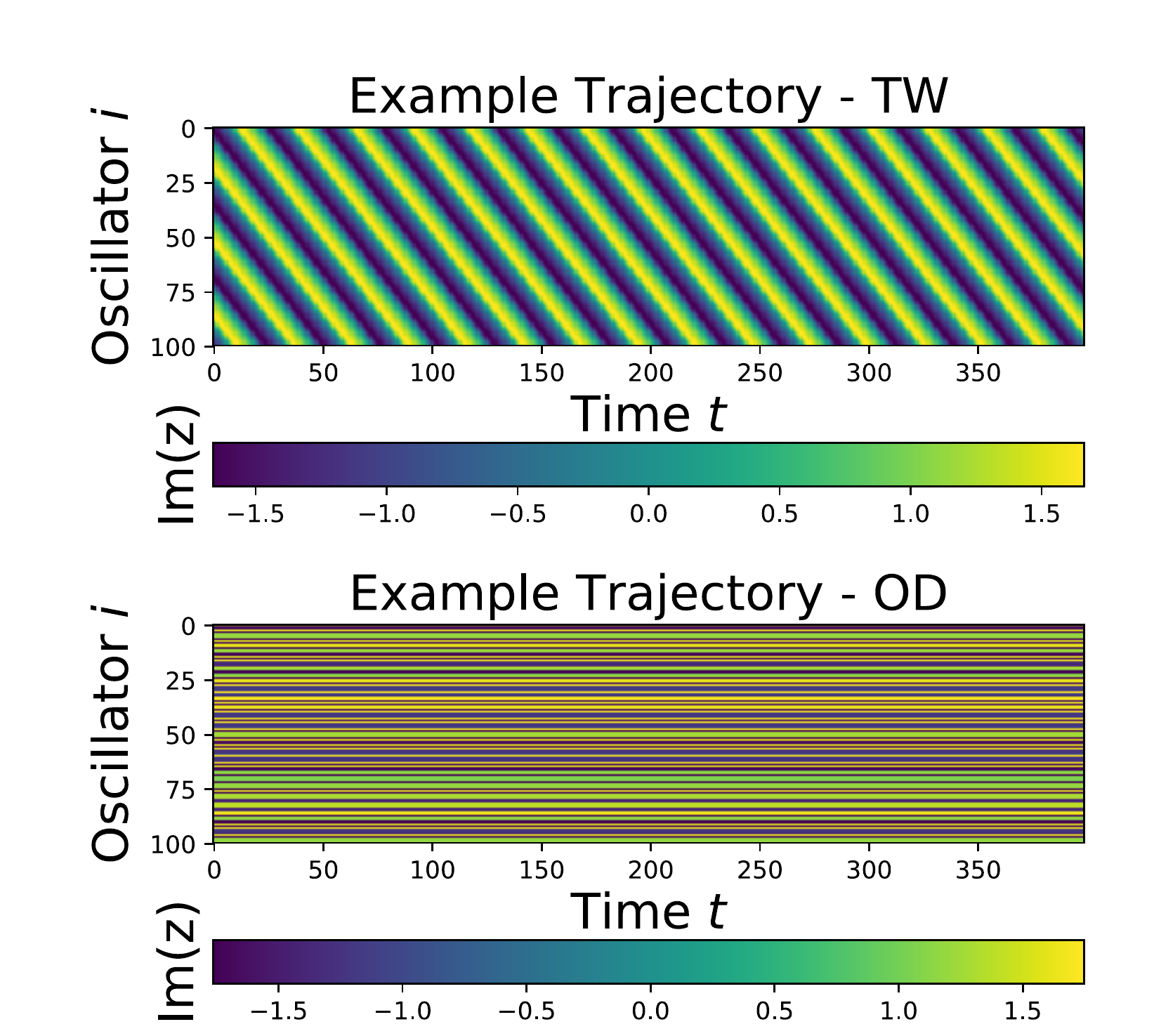}
    \caption{Further analysis on the clusters also shown in Fig. \ref{fig:sl-1d-cm}. (A),(B),(E),(F) are sliding window histograms fits of the denoted measures for trials with parameters within the respective window. (C) and (D) are example trajectories of trials within these clusters.} 
    \label{fig:app-sl-2}
\end{figure}

Additionally to the results presented in Sec. \ref{sec:results}, one can also further inspect the other clusters found by MCBB for the Stuart-Landau systems. This is done in Fig. \ref{fig:app-sl-1} and \ref{fig:app-sl-2}. The Julia package (see Sec. \ref{sec:a:julia} also allows for further other visualizations and inspections of the measures and the clusters. The documentation of the package explains these in more detail.  

\begin{acknowledgments}
The authors thank Jobst Heitzig, Marc Wiedermann and Valerio Lucarini for fruitful discussions about the presented approach. This paper was developed within the scope of the IRTG 1740/TRP 2015/50122-0, funded by the DFG/FAPESP, the Condynet2 project by BmBF FK. 03EK3055A and the DFG project CoCo-Hype KU 837/39-1 / RA 516/13-1.
The authors thank the German Federal Ministry of Education and Research and the Land Brandenburg for supporting this project by providing resources on the high performance
computer system at the Potsdam Institute for Climate Impact Research. 
\end{acknowledgments}

\bibliography{biblio}

\end{document}